\documentclass[aps,prd,twocolumn,nofootinbib,superscriptaddress]{revtex4}
\usepackage[hypertex]{hyperref}
\usepackage{graphicx}

\def\figsize{3 in}
\def\threeone{${\rm SU(3)\! \times\! U(1)}$}
\def\threeonesq{$\left[{\rm SU(3)\! \times\! U(1)}\right]^2$}
\def\twoone{${\rm SU(2)\! \times\! U(1)}$}

\begin{document}

\pagestyle{plain}
\title{Symmetry Breaking Patterns for the Little Higgs from Strong Dynamics}

\author{Puneet Batra} \affiliation{Department of Physics, Columbia
University, 538 W. 120th St., NYC, NY 10027} \author{Z. Chacko}
\affiliation{Department of Physics, University of Maryland, College Park, MD 20742}


\begin{abstract}

We show how the symmetry breaking pattern of the simplest little Higgs model, and that 
of the smallest moose model that incorporates an approximate custodial SU(2), can be 
realized through the condensation of strongly coupled fermions. In each case a 
custodial SU(2) symmetry of the new strong dynamics limits the sizes of corrections 
to precision electroweak observables. In the case of the simplest little Higgs, there 
are no new light states beyond those present in the original model. However, our 
realization of the symmetry breaking pattern of the moose model predicts an additional 
scalar field with mass of order a TeV or higher that has exactly the same quantum 
numbers as the Standard Model Higgs and which decays primarily to third generation 
quarks.

\end{abstract}

 \maketitle


\section{Introduction}

In the Standard Model (SM) the Higgs mass parameter receives quadratically divergent 
radiative corrections from high scales, leading to a hierarchy problem. This suggests 
the existence of new physics at or close to a TeV that cancels these radiative 
corrections, thereby stabilizing the weak scale.

One interesting possibility, first explored in~\cite{GP, KG, Kaplan:1983sm}, is that 
the Higgs mass parameter is protected against radiative corrections because the Higgs 
is the pseudo Nambu-Goldstone boson (pNGB) of an approximate global symmetry. This 
idea has recently experienced a revival in the form of little Higgs 
theories~{\cite{Little1, Little2, Little3, Martin, {Chang:2003un}}}, and also twin 
Higgs theories~{\cite{twin, twinLR}}. These theories stabilize the weak scale against 
radiative corrections up to scales of order 10 TeV.  

It is important to find ultra-violet (UV) completions of these theories that extend 
their validity to scales beyond 10 TeV. Non-supersymmetric weakly coupled UV 
completions of the simplest little Higgs~{\cite{turtle}} and simple group little 
Higgs~\cite{tower} have been found. Supersymmetric UV completions of the little 
Higgs~{\cite{littleSUSY}}, (see also {\cite{littleSUSYold}}), have also been 
constructed, in the context of the supersymmetric little hierarchy problem. There has 
also been work on the difficult problem of finding UV completions of these theories 
where the pattern of symmetry breaking is realized through strong 
dynamics~{\cite{{KLNW}, {Thaler:2005kr}}}. In this case the problem is complicated by 
the requirement that the new strong dynamics preserve a custodial SU(2) symmetry, so 
as not to generate large corrections to precision electroweak observables. An 
alternative approach has been to construct holographic little Higgs models in five 
dimensions~{\cite{holographic}}, which are related to strongly coupled theories in 
four dimensions through the AdS/CFT correspondence.
    
In this paper we show that the characteristic symmetry breaking patterns of two 
well-motivated little Higgs theories can be realized through the condensation of 
strongly coupled fermions. We begin by dynamically realizing the symmetry breaking 
pattern of the simplest little Higgs model. Our construction does not require any 
additional light states beyond those of the original model, and a custodial SU(2) 
symmetry of the new strong dynamics limits the size of corrections to precision 
electroweak observables. We then go on to consider the moose model of Chang and Wacker 
\cite{Chang:2003un}, which is the smallest extension of the minimal moose 
\cite{Little3} which incorporates an approximate custodial SU(2) symmetry, and show 
that this symmetry breaking pattern can also be realized through strong dynamics. In 
the case of this `Next-to-Minimal Moose' (NMM) model, our construction predicts the 
existence of an additional scalar field with mass of order a TeV that has exactly the 
same quantum numbers as the Standard Model Higgs, and which decays primarily to third 
generation quarks.

\section{The Simplest Little Higgs}

Consider first the simplest little Higgs, which has the symmetry breaking 
pattern [\threeone\ $ \! \rightarrow \!$ \twoone]$^2$. The vector 
\threeone\ subgroup, which is gauged, is broken down to the SM gauge group 
\twoone, and 5 of the 10 Nambu-Goldstone bosons (NGBs) are eaten. The 
remaining 5 are actually pNGBs, and they constitute the Higgs of the SM 
and an additional SM singlet.
   
The interactions of the pNGBs are governed by universal low-energy 
theorems which are independent of any specific UV completion. This allows 
us to write down an effective field theory for the pNGBs valid at low 
energies, which takes the form of a non-linear sigma model.

We parameterize the pNGB degrees of freedom as $h^a$ and $\hat{h}^a$. We write
\begin{equation} 
\label{pseudo} 
\phi = \exp(\frac{i}{f} {h}^a t^a) \pmatrix{ 0 \cr  0 \cr f } \; 
\; \; \;
\hat{\phi} = \exp(\frac{i}{\hat{f}} {\hat{h}}^a t^a) \pmatrix{ 0 \cr 0 \cr \hat{f}},
\end{equation} 
where $f$ and $\hat{f}$ are the two independent breaking scales for each \threeone\ 
global symmetry. The 5 matrices $t^a$ span [\threeone/\twoone]:
\begin{eqnarray}
\left\{t^a\right\} &=& \left\{ \pmatrix{ 0 & 0 & 0 \cr
                0 & 0 & i \cr                      
                0 & -i & 0}, \pmatrix{ 0 & 0 &0 \cr
                0 & 0 & -1 \cr
                0& -1 & 0},\  \pmatrix{ 0 & 0 & i \cr
                0 & 0 & 0 \cr
                -i & 0 & 0},\ \right.\nonumber \\
&&\left. \pmatrix{ 0 & 0 & 1 \cr
                0 & 0 & 0 \cr
                1 & 0 & 0}, \sqrt{\frac{2}{3}}\pmatrix{ 1 & 0 & 0 \cr
                0 & 1 & 0 \cr
                0 & 0 & -1} \right\} .
\end{eqnarray}
In general the low energy effective Lagrangian for the fields $h^a, \hat{h}^a$ will 
contain all operators involving $\phi$ and $\hat{\phi}$ consistent with the non-linearly 
realized \threeone\ symmetry, suppressed by the cutoff of the non-linear sigma model, 
which we denote by $\Lambda$. The scale $\Lambda$ is constrained to be less than or of 
order $4 \pi f$, where the upper bound is at strong coupling. The coefficients of these 
operators are determined by the UV physics.

One of the operators allowed by symmetry takes the form
\begin{equation}
\label{problem}
c \; \frac{|\phi^{\dagger} D_{\mu} \phi|^2}{\Lambda^2},
\end{equation}
where the value of the coefficient $c$ depends on the specific
UV completion. This operator violates the approximate custodial SU(2)
symmetry of the Standard Model.  Its effect is
to alter the ratio of the masses of the W and Z gauge bosons from the
Standard Model prediction, and it is therefore very tightly
constrained by experiment.  A potential problem with any UV
completion where this pattern of symmetry breaking is realized through
strong dynamics is that $c$ is expected to be of order $ (4 \pi)^2$,
constraining the cutoff $\Lambda$ to be of order 50 TeV or
higher. Since little Higgs theories can only stabilize the weak scale
up to energies of order 10 TeV or so, this reintroduces
fine-tuning. 

Is there a way that this operator can be suppressed? A linear realization of this 
symmetry breaking pattern is instructive: Consider two scalar fields $\Phi$ and 
$\hat{\Phi}$ which transform as a $3_{-\frac{1}{3}}$ under the \threeone\ gauge 
symmetry. The Lagrangian is
\begin{eqnarray}
\label{linear}
( |D_{\mu} \Phi|^2  + m^2 |\Phi|^2 - \lambda |\Phi|^4 ) \;  + \; ( \Phi \rightarrow 
\hat{\Phi} ). 
\end{eqnarray}
After symmetry breaking one linear combination of the NGBs from $\Phi$ and $\hat{\Phi}$ 
is eaten, while the orthogonal linear combination which contains the SM Higgs survives. 
We denote the VEV of $\Phi$ by $f$ and that of $\hat{\Phi}$ by $\hat{f}$. The pNGB 
fields of the non-linear model are those degrees of freedom which survive after 
integrating out the radial modes of the fields $\Phi$ and $\hat{\Phi}$ in the linear 
model.

The key observation is that in the limit that the gauge symmetry is
turned off this potential, Eq.~(\ref{linear}), has an accidental
SO(6$)^2$ global symmetry which is broken to SO(5$)^2$, and the 10
NGBs can just as well be thought of as arising from this symmetry
breaking pattern. Once we gauge the vector \threeone\ subgroup of the
global symmetry again, 5 of these NGBs are eaten, while the remaining
5 survive in the low energy theory as pNGBs.

Consider now the situation where the breaking pattern SO(6$)^2$ to SO(5$)^2$ is realized 
non-linearly. Since the symmetry breaking pattern now preserves the custodial SU(2) of 
the SM, operators such as Eq.~({\ref{problem}}) are forbidden to leading order. They are 
only generated through loops involving those interactions (gauge and Yukawa couplings) 
that 
violate the custodial SU(2) symmetry, and therefore the coefficient $c$ is at most of 
order one. This allows the scale $\Lambda$ in Eq.~({\ref{problem}}) to be as low as 5 
TeV without conflicting with the constraints from precision electroweak measurements.

We see from this that the first step to finding a strongly coupled UV
completion of the simplest little Higgs is to find a way to break
SO(6$)^2$ to SO(5$)^2$ through strong dynamics. We now explain how
this can be accomplished. Notice that SU(4) is the double-cover of SO(6), and  Sp(4) is the double-cover of SO(5); in both cases the Lie algebras are isomorphic. The problem is therefore to break SU(4$)^2$ to
Sp(4$)^2$ through strong dynamics. Consider an SU(2) gauge theory with
4 fields $ \psi_{\alpha i}$ in the fundamental representation. Here
$\alpha$ is an SU(2) gauge index and $i$ takes values 1 through
4. When the SU(2) gauge theory gets strong we expect a condensate
$\langle \epsilon_{\alpha \beta} \psi_{\alpha i} \psi_{\beta j}
\rangle \propto J_{ij}$ to form along the gauge singlet
direction. This is antisymmetric in the indices $i$ and $j$, thereby
breaking the SU(4) global symmetry down to Sp(4).\footnote{More generally,
any condensing Sp(2N) with four fundamental flavors of fermions can be
used to break SU(4) to Sp(4) \cite{Kaplan:1983sm}.}  In order to break
SU(4$)^2$ to Sp(4$)^2$ we merely begin with two copies of the
SU(2) gauge theory---each with 4 fields in the fundamental
representation. We label the two copies of the $\psi$'s as
$\psi_{\alpha i}$ and $\hat{\psi}_{\alpha i}$, and the corresponding
condensates $J_{ij}$ and $\hat{J}_{ij}$.

\begin{figure}[t]
\includegraphics[width=\figsize]{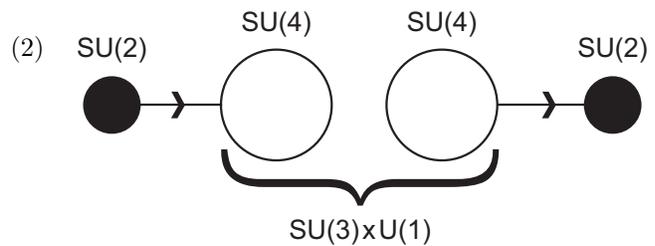}
\caption{A UV completion of the simplest little Higgs. The flavor SU(4$)^2$ symmetry has 
gauged subgroup \threeone. The curly braces indicate that only the diagonal subgroup of 
SU(4$)^2$ is gauged. The black filled-in circle is a condensing SU(2) group. Each link 
field is a collection of four SU(2) fundamentals that transforms like a fundamental 
(anti-fundamental) of SU(4).}
\label{fig:su3}
\end{figure}

We are free to weakly gauge a vector \threeone\ subgroup of the
SU(4$)^2$ global symmetry, as shown in Figure \ref{fig:su3}. Without
loss of generality we may take the indices $i =1,2,3$ to be SU(3)
gauge indices, while $i = 4$ is not an SU(3) gauge index. The SU(3)
gauge symmetry is anomaly free provided the fields in $\psi$ transform
in the fundamental representation and the fields in $\hat{\psi}$ in
the anti-fundamental. We choose the charges of the U(1) such that the
fields in $\psi$ transforming as the fundamental of SU(3) have charge
+1/6, while the SU(3) singlet has charge -1/2. The anti-fundamental of
SU(3) in $\hat{\psi}$ has U(1) charge -1/6 and the SU(3) singlet
charge +1/2. After condensation this \threeone\ gauge symmetry is
broken to the \twoone\ of the SM.

The low energy effective theory can once again be described by a non-linear sigma 
model.
The matrix $J_{ij}$ has the structure
\begin{equation}
J  =  \pmatrix{ i \sigma^2 & 0\cr
                         0& i \sigma^2}.
\end{equation}
The matrix $J$ is left invariant under Sp(4) rotations, but all other generators of 
SU(4) are broken. We parameterize the five resulting low energy degrees of freedom as 
$\pi^a$. Define
\begin{equation}
A =  f {\rm exp} ( \frac{i}{f} \pi^a T^a ) \; J \; {\rm exp} ( \frac{i}{f} \pi^a T^a 
)^T.
\end{equation}
Here the 5 matrices $T^a$ are the generators of SU(4)/Sp(4). 
\begin{equation}
\left\{ T^a \right\} =  \left\{ \pmatrix{ 0 & i \sigma^a \cr
                                       - i \sigma^a & 0}, \pmatrix{ 0 & I \cr
                                        I & 0}, \pmatrix{ I & 0 \cr
                                        0 & -I}\right\}.
\label{sp4gen}
\end{equation}
The matrix $\hat{J_{ij}}$  has exactly the same form as the matrix
$J_{ij}$, and we parameterize the corresponding low energy degrees of
freedom by $\hat{\pi}^a$, collected in a matrix $\hat{A}$.

By choosing to keep only the \threeone\ subgroup of each SU(4) symmetry manifest, we 
can immediately carry over all results of the original simplest little Higgs model to 
this new construction. To see this explicitly, consider the decomposition of SU(4) 
into its \threeone\ subgroup. We can identify the $\pi^i$'s and $\hat{\pi}^i$'s with 
the $h^i$'s and $\hat{h}^i$'s up to constants of proportionality, since they have 
exactly the same transformation properties under the nonlinearly realized \threeone:
\begin{equation}
A= \pmatrix{0  & \phi_3^{\star} & -\phi_2^{\star} & \phi_1 \cr
-\phi_3^{\star} &0 &  \phi_1^{\star} & \phi_2 \cr
\phi_2^{\star} & -\phi_1^{\star} & 0 & \phi_3\cr-\phi_1 & -\phi_2 & -\phi_3 & 0}.
\end{equation}
In the absence of any explicit symmetry breaking, the low energy effective Lagrangian 
for the pNGBs consists of all operators involving $A$ and $\hat{A}$ consistent with the 
non-linearly realized SU(4$)^2$ symmetry. Equivalently, the low energy effective 
Lagrangian consists of all possible operators involving $\phi$ and $\hat{\phi}$ 
consistent with the nonlinearly realized \threeonesq\ symmetry, but with additional 
relations among the coefficients of the various terms enforced by the larger SU(4$)^2$ 
global symmetry. In particular, dangerous operators which violate the custodial SU(2) 
symmetry, such as that in Eq.~({\ref{problem}}), are forbidden at leading order.

Any potential for the pNGBs can arise only from those interactions that explicitly 
violate the ${\rm SU(4)^2}$ global symmetry, in particular the \threeone\  gauge symmetry and 
the Yukawa couplings. In the low-energy effective Lagrangian these interactions can be 
written down explicitly in terms of the fields $\phi$ and $\hat{\phi}$, exactly as in 
the original model. The SM fermions can be embedded, anomaly-free, under \threeone, by 
promoting SM doublets of SU(2) to triplets or anti-triplets as required 
\cite{Martin},\cite{Kong:2003tf}. The Yukawa interactions of the low energy effective 
theory arise from non-renormalizable interactions between the SM fermions and the 
fermions that condense to form the Higgs. Consider, for concreteness, the top Yukawa 
coupling.
The third-generation left-handed quark doublet, $q$, is enlarged to an SU(3) 
triplet with U(1) charge 1/3, $Q = (q, U)$. This triplet couples to two $SU(3)$ 
singlets $U_1^c$ and $U_2^c$ through the following terms:
\begin{equation}
\label{NRYukawa}
\frac{\lambda_1}{4 \pi f^2} \left(\psi_i \psi_4 \right)^{\dagger} Q_i U_1^c + 
\frac{\lambda_2}{4 \pi f^2} \left(\hat{\psi^i} \hat{\psi^4}\right) Q_i U_c^2,
\end{equation} 
where only the SU(3) gauge indices are shown explicitly. This leads to the 
 couplings below in the low energy effective theory after the SU(2) groups get 
strong:
\begin{equation}
{y_1} \; \phi_{i} \; Q_i U_1^c +
{y_2} \; \hat{\phi}_{i} \; Q_i U_2^c.
\end{equation}
These interactions, which are familiar from the original model, give rise to the top 
Yukawa coupling while eliminating the one-loop quadratic divergences from the top loop. 
We leave the important question of the UV origin of the higher dimensional operators 
in Eq.(\ref{NRYukawa}) for future work.

\begin{figure}[t]
\includegraphics[width=\figsize]{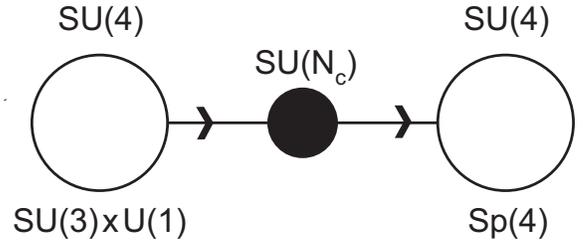}
\caption{An alternate moose for the ${\rm SO(6) \! \rightarrow \! SO(5)}$ symmetry 
breaking pattern. The moose has an ${\rm SU(4)^2}$ flavor symmetry, with gauged 
subgroup ${\rm Sp(4) \! \times \! SU(3) \! \times \! U(1)}$.  One condensing ${\rm 
SU(N_c)}$ group is also present, and each link field is a bifundamental fermion under 
the neighboring global symmetries.}
\label{fig:su3alt}
\end{figure}

It is interesting to note that there is another configuration which also generates 
exactly the symmetry breaking pattern of the simplest little Higgs with a custodial 
SU(2) symmetry. As pointed out in \cite{Thaler:2005kr}, the global symmetry breaking 
pattern ${\rm G \! \rightarrow \! H}$ with a subgroup F of G gauged has the same low 
energy dynamics as a two-site nonlinear sigma model with global symmetry breaking 
pattern ${\rm G^2 \! \rightarrow \! G}$ and gauged subgroup ${\rm H\! \times \! F}$, 
in the limit that the gauge coupling constant of H is large. This two-site model can 
in principle be UV completed, provided the ${\rm G^2 \! \rightarrow \! G}$ breaking 
pattern is a generalization of QCD-type confining dynamics.

Since the pattern ${\rm SO(6) \!  \rightarrow \! SO(5)} \simeq {\rm SU(4) 
\! \rightarrow \! Sp(4)}$, we can use this technique to realize symmetry breaking in 
the simplest little Higgs. The appropriate construction is shown in Figure 
(\ref{fig:su3alt}); to replicate exactly the simplest little Higgs we simply repeat 
this construction twice and gauge the same \threeone \ symmetry in each case.

\section{The Next-To-Minimal Moose}

The NMM model \cite{Chang:2003un} is built around the symmetry breaking pattern ${\rm 
SO(5)}^2 \! \rightarrow \! {\rm SO(5)}$. The same pattern also arises in several other 
little Higgs models {\cite{SO5}}. We now show that the symmetry breaking pattern ${\rm 
Sp(4)^2 \! \rightarrow \! Sp(4)}$, which is equivalent to ${\rm SO(5)^2 \! \rightarrow 
\! SO(5)}$, can be realized through strong dynamics. Consider an SU(${\rm N_c}$) gauge 
group, with a set of four fermions, $\chi_{\alpha i}$, in the fundamental 
representation. Here $\alpha$ represents an SU(${\rm N_c}$) gauge index and $i$ 
labels the fermions from 1 through 4. We also add a set of four fermions in the 
conjugate representation $\hat{\chi}^{\alpha}_i$. When the SU(${\rm N_c}$) theory gets 
strong, a condensate $\langle \hat{\chi}^{\alpha}_i 
\chi_{\alpha j} \rangle \propto \delta_{ij}$ forms and breaks the ${\rm SU(4)_ L \! \times \! SU(4)_R}$ 
flavor symmetry to the diagonal SU(4). We label the 15 resulting NGBs that are 
produced by $\pi^a$, and define

\begin{equation}
X = f {\rm exp}\left({2 i}\pi^a T^a/f \right),
\end{equation}
where the matrices $T^a$ are generators of SU(4).

We also add to the theory a non-renormalizable term 
\begin{equation}
\frac{m^2}{\left(4 \pi f^2 \right)^2} {\rm Tr} \left[ \left( \chi \hat{\chi}\right) J 
\left( \chi \hat{\chi}\right)^T J \right] \sim m^2 {\rm Tr} \left[ X J X^T J \right]
\end{equation}
which is allowed by the gauge symmetries.
The effect of this term is to explicitly break the global ${\rm SU(4)}^2$ symmetry to 
${\rm Sp(4)}^2$, thereby giving a mass of order $m$ to 5 of the 15 NGBs. With the 
addition of this term the pattern of global symmetry breaking is in fact ${\rm 
Sp(4)^2 \! \rightarrow \! Sp(4)}$, which accounts for the 10 surviving NGBs. The 
unbroken global symmetry, the diagonal Sp(4), contains the custodial SU(2) symmetry we 
desire. 

\begin{figure}[t]
\includegraphics[width=\figsize]{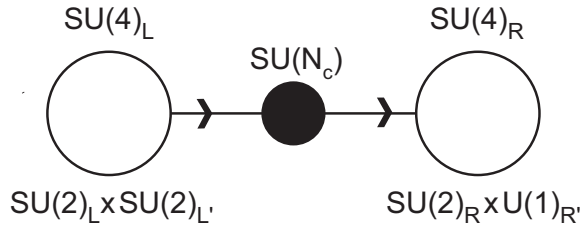}
\caption{A UV completion for the minimal moose. The theory has a global ${\rm SU(4)_L 
\! \times \! SU(4)_R}$ flavor symmetry with the indicated (written along the bottom) 
gauged subgroup. There is also a condensing SU(${\rm N_c}$) gauge group. Each link 
represents a bifundamental fermion under the indicated symmetries.}
\label{fig:nmm}
\end{figure}

To reproduce the NMM gauge symmetry breaking pattern, we weakly gauge an ${\rm SU(2)_L 
\! \times \! SU(2)_{L'}}$ subgroup of the ${\rm SU(4)_L}$ symmetry, and an ${\rm 
SU(2)_R \! \times \!  U(1)_{R'}}$ symmetry of the ${\rm SU(4)_R}$ global symmetry, as 
shown in Figure \ref{fig:nmm}. Here ${\rm U(1)_{R'}}$ is the diagonal generator of the other SU(2$)_{\rm R'}$ subgroup contained in ${\rm SU(4)_R}$. The gauge symmetry is broken 
to the diagonal \twoone\ of the SM and 6 NGBs are eaten.  The remaining 4 light pNGBs 
constitute a single complex Higgs doublet. The custodial SU(2) symmetry is preserved by 
the kinetic terms up to small corrections arising from integrating out the states of 
mass $m$. Provided $m$ is larger than or of order a TeV, these are under control. This 
is in contrast to the original minimal moose where corrections to the precision 
electroweak observables arising from the kinetic terms are in general very large
\cite{Rakhi}. 

It is now straightforward to reproduce the complete symmetry breaking pattern of the 
NMM model by repeating the breaking pattern ${\rm SO(5)^2 \! \rightarrow \!  SO(5)}$ 
multiple times. Here we limit ourselves to constructing a model with just a 
single link that reproduces the main features of the NMM model.

The fermion sector is straightforward and anomaly-free. Each SM SU(2) doublet, $q$ or $l$, 
transforms under  ${\rm SU(2)_L}$ and emerges from a fundamental of ${\rm SU(4)_L}$, 
which we denote by $\hat{q}$ or $\hat{l}$. A second doublet, $q'$ or $l'$, which 
transforms under ${\rm SU(2)_{L'}}$ fills out $\hat{q}$ or $\hat{l}$. The fields $\hat{q}$ 
and $\hat{l}$ carry ${\rm U(1)_{R'}}$ charges equal to the SM hypercharges of $q$ and $l$ 
respectively. In addition, corresponding to each $q'$ or $l'$ is a field $q'^c$ or 
$l'^c$ that is vector-like with respect to it. The SM SU(2) singlet fields, $U^c, D^c$ 
and $E^c$, only carry ${\rm U(1)_{R'}}$ gauge quantum numbers with charges again equal 
the 
corresponding SM 
hypercharges. The quark Yukawa couplings emerge from
\begin{equation}
\label{topYuk}
{\cal L} \supset \frac{1}{ 4 \pi f^2}
\pmatrix{q \cr q'}\cdot \left( \chi \hat{\chi}\right) \pmatrix{
0\cr 0 \cr \lambda_U U^c \cr \lambda_D D^c} + \lambda' f q' q'^c.
\end{equation}
One linear combination of $U^c$ and $U'^c \subset q'^c$ acquires a mass of order $f$, 
while the other linear combination is the usual SM singlet, $u^c$. A similar story 
holds for the down-type quarks. The invariance of the first term under SU(4$)_{\rm L}$ guarantees the cancellation of one loop quadratic divergences
from the quark sector. 
The generalization to leptons is straightforward.
 
The states with mass $m$ are not present in the original NMM model, but constitute a 
firm prediction of our construction. In particular, they will be present in any UV 
completion that realizes the pattern of symmetry breaking in the same manner. These 
states are composed of a field with the same SM quantum numbers as the Higgs that decays 
predominantly into third generation quarks, and a SM singlet.

In conclusion, we have shown how to obtain the symmetry breaking patterns of the 
simplest little Higgs and the smallest moose with an approximate custodial SU(2) symmetry 
from strong dynamics. Our hope is that this will open the door to the construction of 
completely realistic, UV completions of the little Higgs from strong dynamics.  \\

\noindent
{\bf Acknowledgments --} The authors would like to thank Spencer
Chang, Markus Luty and Eduardo Ponton for useful discussions. PB is
supported by the DOE under contract DE-FG02-92ER.  ZC is supported by
the NSF under grant PHY-0408954.

\end{document}